# The Impact of Retail Investors' Sentiment

# on Conditional Volatility of Stocks and Bonds[i]

Elroi Hadad[ii]

Shamoon College of Engineering (SCE)

Haim Kedar-Levy[iii]

Ben-Gurion University of the Negev

---

[i] We are thankful to Shmuel Hauser, the Ono Academic College, and Bizportal.co.il for granting us the opportunity to access and explore their unique sentiment indexes. Financial aid by the Ono Research Fund is gratefully acknowledged. Should any errors remain in the text, they are due to the authors.

[ii] Shamoon Collage of Engineering (SCE), Beer-Sheva, Israel.

[iii] The Guilford Glazer Faculty of Business and Management, Ben-Gurion University of the Negev, Israel. P.O.B. 653, Beer Sheva 84105, Israel. hlevy@bgu.ac.il Tel. (972) 8-6472-569, Fax (972) 8-6477 697. Corresponding author.

# The Impact of Retail Investors' Sentiment

# on Conditional Volatility of Stocks and Bonds

## Abstract

We measure bond and stock conditional return volatility as a function of changes in sentiment, proxied by six indicators from the Tel Aviv Stock Exchange. We find that changes in sentiment affect conditional volatilities at different magnitudes and often in an opposite manner in the two markets, subject to market states. We are the first to measure bonds' conditional volatility of retail investors' sentiment thanks to a unique dataset of corporate bond returns from a limit-order-book with highly active retail traders. This market structure differs from the prevalent OTC platforms, where institutional investors are active yet less prone to sentiment.

# 1. Introduction

Theories of behavioral finance argue that asset prices may be affected by behavioral and psychological patterns, such as overconfidence, optimism, wishful thinking, etc., of individual (retail) market investors who are subject to shifts in sentiment (see, among others, Kyle, 1985; Black, 1986; Shleifer and Summers, 1990; Brown and Cliff, 2004).[1] The effect of changes in sentiment on stock returns is highly evident in the literature (Brown and Cliff, 2004; Baker and Wurgler, 2006; Baker and Wurgler, 2007), and is known to positively affect the volatility of stock returns, suggesting that retail investors may be viewed as noise traders (see also Kumar and Lee 2006; Kaniel, Saar and Titman, 2007; Barber and Odean, 2008; Foucault, Sraer and Thesmar, 2011). However, while studies show strong evidence of spillover effects between equity and bond returns (Downing, Underwood and Xing, 2009; Hong, Lin and Wu, 2012), the effect of investor sentiment on bond returns is discussed in the literature only to a limited extent. The few exceptions study the sentiment effect in US or European corporate bond markets, where bonds are traded in decentralized, dealer intermediated, Over-The-Counter (OTC) markets.[2] These OTC markets are mostly characterized by a low transparency and high transaction costs, which limit the participation rate of individual investors in these markets (Edwards, Harris and Piwowar, 2007), and, intuitively, limit the effect of sentiment on bond returns.

---

[1] E.g., Barber and Odean (2008) and Karlsson, Loewenstein and Seppi (2009) show that individual sentiment traders tend to buy more aggressively than to sell attention-grabbing stocks (i.e., stocks in the news, stocks with extreme one day returns, etc.), hence have a much greater effect on stock prices during high sentiment periods; Yu and Yuan (2011) show that a low tradeoff between stock market expected returns and risk during high sentiment periods is explained by higher participation rate of sentiment-driven traders.

[2] Nayak (2010) and Bethke, Gehde-Trapp and Kempf (2017) show that US corporate bonds yield spreads co-vary with sentiment in in a similar way to stocks (i.e., underpriced during pessimistic periods and overpriced when optimism reigns). Spyrou (2013) documents the investor sentiment effect on corporate bonds yield spreads for a sample of European markets, in the light of the recent subprime crisis.

In this paper we hypothesize that a corporate bond market in which retail trading activity is relatively high would be more susceptive to changes in sentiment, and hence might appear riskier than an equivalent OTC market (Foucault et al., 2011). To address this hypothesis, we harness a unique dataset of the Tel Aviv Stock Exchange (TASE) where retail investors are highly active in corporate bonds trading as the trading platform is a continuous limit order book (LOB). The test has broad implications for investors and issuers since the volatility of returns is a key variable in various aspects of finance, such as asset pricing, risk management, and liquidity.[3]

As noted, the corporate bonds market in TASE is identical to the open LOB platform that operates for stocks, in TASE and in most exchanges worldwide. Prior research indeed documented that corporate bond trading volume in TASE is relatively high with a substantial presence of retail-size traders (Abudy and Wohl, 2017). These properties make it highly relevant for studying how changes in sentiment of retail-size investors affect the volatility of bond returns.[4,5]

Yet, whatever the findings would be with respect to the corporate bonds market, they would carry little meaning unless compared with a more studied and well documented platform, i.e., that of the stock market. Because both markets run on an identical trading platform and investor base, with contemporaneous changes in sentiment, comparing the impact of sentiment between the corporate bonds and the local stock market would allow us to gauge in comparative terms the implications of sentiment. We therefore analyze the level of sentiment effects in each market separately, and then discuss and compare the two. Further, since sentiment is more evident in a bearish than bullish market states (e.g., Lee, Jiang and Indro, 2002), we test the magnitude of the sentiment effect under different market states.

---

[3] Amihud and Mendelson (2006) show that bond liquidity bears risks to investors, and hence liquidity affects the bond's yield-to maturity and the issuers' cost of capital.

[4] The 2019 annual statistics from the World Federation of Exchanges (WFE) shows that the number of trades in bonds in the Tel-Aviv Stock Exchange (TASE) exceeded 10.98 million trades with total value traded above 240 billion USD.

[5] Abudy and Wohl (2017) note that retail investors are highly active in the Israeli corporate bonds market, and many of the short-term traders are small trading firms and individual traders, which improves the liquidity of corporate and government bonds. Gur-Gershgoren et al., (2020) find this market "deep" in the context of IAS-19 accounting standard. Kedar-Levy et al. (2020) verify this result following the shock to bond prices due to the Covid-19 outbreak.

Using the EGARCH (1,1) model (Nelson, 1991), we analyze return volatilities of major bond and stock indices at the TASE as a function of changes in several sentiment proxies. Our main findings are that the conditional volatilities of both stock and bond returns are affected by changes in retail sentiment, and that the magnitude of this effect differs between these markets and is subject to bullish or bearish market conditions. We show that under normal market conditions, when sentiment is estimated to be relatively high, changes in momentum-based sentiment proxies largely explain changes in the conditional volatility of the bond and stock index returns, but hardly explain the variability of these indexes in times of financial turmoil. We also find that in conditions of very low sentiment, e.g., during the financial crisis of 2008, a change in volatility-based sentiment proxies has a larger effect on the conditional volatility of bond returns than in times of normal market conditions, a pattern which is reversed when considering stock returns. These results imply that given a high presence of retail investors, changes in sentiment also affect the volatilities of bond returns. This pattern may be important for institutional traders and regulators who may interpret bond market volatility in a rational manner and calm the public in times of stress. Our findings further imply that different sentiment proxies are more informative than others in different market states and they differ between the two markets.

The rest of the paper is organized as follows: Section 2 discusses the theoretical effects of investor sentiment on asset return volatility. Section 3 presents the unique dataset, and Section 4 elaborates on the methodology we implemented. Section 5 presents the results separately for stocks and for bonds, and Section 6 summarizes and concludes the paper.

## 2. Theoretical effects of retail sentiment on volatility

The motivation for this study stems from two assumptions that relate investor sentiment to asset prices as laid out in Baker and Wurgler (2007): (1) An investor's sentiment, broadly defined as an investor's belief about future cash flows and investment risks, may not be justified by fundamental news or facts, and (2) it is costly and risky to bet against sentimental investors, and thus rational investors, or arbitragers, are not as aggressive in forcing prices back to risk-based fundamentals (i.e., there are limits to arbitrage).[6] These assumptions suggest that the *level* of the effect of investor sentiment on assets prices is also given by the participation rate of retail traders ("noise traders") in the markets. On that point, Foucault et al. (2011) have shown that the presence of retail trading activity in markets also has a positive and significant effect on the volatility of stock returns. This effect has broad implications since if noise traders have a high impact on the volatility of returns, they impose a higher risk on rational arbitrageurs, and hence have a greater impact on assets prices. In that sense, a high presence of retail trading in corporate bond markets may also affect the volatility of corporate bond returns, thus may increase the cost of debt due to sentiment.

In order to study how sentiment may affect volatility, we follow the strand of the literature which relates sentiment to the conditional volatility of stock returns. Lee et al. (2002) show that changes in investor sentiment are inversely correlated with the conditional volatility of US stock market indexes (Dow Jones Industrial Averages (DJIA), the S&P500 and the NASDAQ), i.e. when investors become bullish the conditional volatility goes down and vice versa. In that sense, Yu and Yuan (2011) also show evidence of a low (high) positive tradeoff between the mean and variance of US stock returns during high (low) sentiment periods, indicating that the greater participation rate of sentiment-driven investors during high-sentiment periods exerts a greater influence on stock prices.

---

[6] Baker and Wrugler (2007) explain that price anomalies may form when sentiment investors over (under) estimate return and underestimate (overestimate) risk. Hence, investing more in the riskier (safer) asset may cause a mispricing of the asset relative to its risk-based fundamental.

Verma and Verma (2007) study the asymmetric effect of changes in sentiment, for both institutional and retail investors, on the conditional volatility of major US stock return indexes by employing the E-GARCH (1,1) model. Verma and Verma (2007) find a significant and negative effect of changes in sentiment of irrational retail-size investors on the volatility of returns (and an insignificant effect for the rational (institutional) investors), suggesting that the presence of retail investors is a significant determinant of stock volatilities. Furthermore, Verma and Verma (2007) show the greater effect of irrational bullish over bearish investors on the volatility of stock returns, concluding that changes in irrational investor sentiment are a critical factor for modeling market volatility. These findings are likewise present in the few studies which relate the sentiment effect to the volatility of European stocks markets.[7]

Following Verma and Verma (2007), we propose to model the asymmetric effect of changes in sentiment, proxied by several trading activity indicators, on the conditional volatilities of returns in the Israeli stock and bond markets.[8] In general, the literature uses a variety of market trading activity indicators as imperfect proxies of sentiment in markets, as these indicators may signal bullish or bearish trends in markets. Thus, Brown and Cliff (2004) use the ratio of new highs to new lows (HI/LO), designed to capture the relative strength of the market. Baker and Stein (2004), Baker and Wurgler (2004, 2007), and Kaniel et al. (2007) use the trading volume (VOL) and the

---

[7] Uygur and Taş (2014) similarly document that changes in daily trading volumes, a proxy for investor sentiment, have a significant effect on the conditional volatility of daily returns measured in the Istanbul stock market.

[8] The sentiment proxies we use were suggested by Prof. Shmuel Hauser of Ono Academic College and Ben Gurion University and were computed and published by the financial website Bizportal.co.il.

Turnover Ratio (TURN), which capture relative liquidity in markets, as proxies for investor sentiment. Dennis and Mayhew (2002) and Brown and Cliff (2004) suggest the Put-Call Ratio (PCR), which measures the ratio of the volume of put and call options contracts, as an indirect sentiment indicator, proxying for the expected beliefs of price drops vs. price increases. Whaley (2000) proposes a Volatility Index (VIX), which many consider a "fear index" to measure the implied volatility of S&P100 stock index options.[9]

Based on the above-mentioned measures of sentiment, we use six measures of the local TASE market, and in particular the continuous trade in the bond market. While we elaborate on each of our six sentiment proxies in the text, we note here that they are made of two groups, one for stocks and the second for bonds.

## 3. Data

This section describes our data set. Our data on investor sentiment in the TASE is drawn from two sources. First, we use TASE equity and debt market indices as published in the TASE website. To quantify investor sentiment in the Israeli equity and debt markets, we use two major indices of asset prices in these markets. For the equity market, we use the TA-35 index, which is a value-weighted index of the 35 largest market capitalization companies listed on the TASE. For the debt market, we use the Tel-Bond-20 index, which consists of 20 corporate bonds with the highest market capitalization of issues among all the bonds traded in the local debt market. All are fixed-interest and linked to the Consumer Price Index.

---

[9] Other sentiment proxies are the dividend premium (DIV) (Baker and Wurgler, 2004), measured as the difference between the average market-to-book value ratios of dividend payers and nonpayers; the close-end-fund discount (CEF) (Lee, Shleifer and Thaler, 1991; Neal and Wheatly, 1998), measured as the difference between the net asset value of funds which issued a fixed number of shares and their respective market prices; and Initial Public Offering (IPO) measures like the first-day-return and IPO-Volume (Ljungqvist, Nanda and Singh, 2006; Cornelli, Goldreich and Ljungqvist, 2006), which reflect investor sentiment in the pricing of stock IPOs.

Second, we use the Bizportal's market sentiment indicators as proxies for investor sentiment in the TASE.[10] These sentiment proxies are based on several trading activity measures of the Israeli stock and bonds markets.

### 3.1. *Stock* market sentiment indicators

For the stock market, we use three indicators that proxy for sentiment in the market. The first indicator is the Stock Market Momentum Index (SMMI), designed to capture a positive (or negative) momentum in the market. For this indicator we augment Daniel, Hirshleifer and Subrahmanyam (1998) who show that the price-momentum effect is generated by investor overconfidence and self-attribution bias, i.e. a positive price momentum in the short run may indicate positive sentiment in the market.[11] We measure the *SMMI* as the ratio between the last 5-day's moving average of the TA-35 index and its moving average in the last 250 trading days. A positive *SMMI* may indicate a positive momentum in the short run and may also be interpreted as a bullish signal in the market.

The second indicator is the Stock Market Sentiment Index (SMSI), which captures a positive (or negative) sentiment in the market based on the Put-Call-Ratio measure proposed by Dennis and Mayhew (2002) and Brown and Cliff (2004). The *SMSI* indicator is calculated as the ratio of the volume of put and call options contracts written on the underlying TA-35. Here a rise in *SMSI* may be interpreted as a bearish signal and as reflecting negative sentiment in the market.

---

[11] There is an ongoing debate in the literature whether time-series momentum stems from behavioral or rational trading patterns by different investor types. For example, Daniel et. al (1998), Barberis, Shleifer and Vishny (1998) and others advocate for behavioral effects, such as representativeness and conservatism or overconfidence and self-attribution. For their part Kedar-Levy (2013) show that such trading patterns may be rational, and Moskowitz, Ooi and Pedersen (2012) show that time-series momentum is unrelated to sentiment.

The third sentiment indicator is the Stock Volatility Index (SVIX), and is based on the implied volatility index of Whaley (2000). The *SVIX* is measured as daily observations of implied volatilities of at-the-money put and call options of the TA-35 index. A rise in *SVIX* may indicate higher anticipated volatility, which may be interpreted as a bearish signal and as negative sentiment.

### 3.2. *Bonds* market sentiment indicators

For the bonds market, we define three sentiment proxies that capture momentum and volatility in the market. The first indicator is the Bonds Market Momentum Index (BMMI), which captures momentum in the short run. The *BMMI* is calculated as the ratio between the 5 days moving average of the Tel-Bond-20 index and its moving average in the last 250 trading days. Similar to *SMMI*, a positive *BMMI* may indicate for a positive sentiment in the bonds market.

The second indicator is the Bond Market Stability Index (BMSI), which is designed to capture the relative risk in the market. The *BMSI* is calculated as the volatility of daily returns of the Tel-Bond-20 index. A rise in *BMMI* may indicate higher risk in the bonds market, which may be interpreted as a bullish signal and as negative sentiment in the market.

The third indicator is the Default Risk Index (DRI), designed to capture the relative risk of default in the bonds market. We argue that bonds with a yield-to-maturity above 8% reflect a higher risk of default, and hence we define the *DRI* as the ratio of the market value of all bonds with respective yield-to-maturity above 8% to the market value of all traded bonds. We argue that a rise in *DRI* may indicate a strong *flight-to-quality* in the market, and hence may be interpreted as a bullish signal and as negative sentiment.

Our sample consists of daily values of the TA-35 and Tel-Bond-20 indexes, along with their respective market sentiment indicators. The sample period runs from January/2000 to March/2019. The dataset begins on the date that the TA-35 and Tel-Bond-20 indexes were launched by the exchange.[12] Finally, we conduct our regression and EGARCH (1,1) analyses on simple daily returns data, computed from adjusted TA-35 and Tel-Bond-20 index levels.[13] In the next sections we show the regression and the EGARCH (1,1) fitting model methodology we used for quantifying the investor sentiment effect in the TASE.

## 4. Methodology

We study how changes in investor sentiment affect the conditional volatility of returns of stocks and bonds in the TASE. We apply an EGARCH model on the returns of TA-35 and Tel-Bond-20 indexes in order to explore the asymmetric effect of the sentiment indicators on the indexes' conditional volatilities.

Our point of departure is a random-walk model aimed at determining the extent to which past returns explain variability in daily returns of TA-35 and Tel-Bond-20 indexes. This test is consistent with the weak-form market efficiency hypothesis whereby past returns of the *k*-type index capture all relevant economic information that affects index values. Specifically, we apply a time-series regression of the form

$$R_{k,t} = \beta_0 + \beta_1 R_{k,t-1} + \varepsilon_{k,t}, \tag{1}$$

where $R_{k,t}$ is the daily return of the *k*-type index (*k*={TA-35; Tel-Bond-20}), *t* represents time and $\varepsilon_{k,t}$ is the error term of (1).

---

[12] It should be noted that the data for the SMSI indicator has some missing information and that the data available for the Tel-Bond-20 returns and its respective market sentiment proxies starts from October 2, 2008.

[13]  We used continuously compounded returns yet found no noticeable differences.

Next, we model the unexplained portion of the daily variations in the $k$-type index, which results in other non-fundamental factors, e.g., investor sentiment and time-series momentum. We model the squared residuals of the $k$-type regression by using the proposed investor sentiment proxy in the following regression,

$$\left(R_{k,t} - \hat{R}_{k,t}\right)^2 = \beta_0 + \beta_1 SENT_{k,i,t} + \varepsilon_{k,t}, \tag{2}$$

where $\hat{R}_{k,t}$ is the fitted value of the $k$-type regression in equation (1), $SENT_{k,i,t}$ is the $i$-type investor sentiment proxy related to the $k$-type index, and $\varepsilon_{k,t}$ is the residual of (2). Hence, for each $k$-type index, we run separate regressions of the squared deviations obtained from equation (1) and the respective $i$-type investor sentiment proxy, namely $i=\{$SMMI, SMSI, SVIX$\}$ for the TA-35 index and $i=\{$BMMI, BMSI, DRI$\}$ for the Tel-Bond-20 index.

The null hypothesis for the $i$-type investor sentiment proxy is $H0$: $\beta_i = 0$; rejection of the null will indicate that the investor sentiment proxy explains variations in the $k$-type market index. Further, in order to answer which of the $i$-type investor sentiment indices better explains variations in the market, we compare the adjusted $R^2$ results from equation (2).

Additionally, we study the effect of the change in sentiment proxies on the conditional volatility of the daily returns in both stocks and bonds indexes. Following Verma and Verma (2007) and Uygur and Taş (2014), we suggest employing an EGARCH (1,1) model (Nelson, 1991) over the daily returns of the TA-35 and Tel-Bond-20 indexes in order to explore the asymmetric effect of investor sentiment on the conditional volatility of the indexes. The motivation for choosing the EGARCH model rather than other models of volatility (e.g., GARCH, GARCH in–mean etc.) is that the EGARCH model can identify asymmetric effects on conditional volatility (Nelson, 1991), i.e. whether a negative shock leads to subsequent conditional variance that differs from a positive shock. Further, the EGARCH model specifies the conditional variance in logarithmic form and hence avoids negative variance estimation; it was also found to be the most successful model for forecasting TASE indexes returns from among the various GARCH models (Alberg, Shalit and Yosef, 2008).[14]

We use the daily returns of the TA-35 and Tel-Bond-20 index as dependent variables in the following EGARCH(1,1) model,

$$r_{k,t} = \mu_k + \varepsilon_{k,t}, \tag{3}$$

where $r_{k,t}$ is the return of the $k$-type index, $\mu$ is a constant term relevant for index $k$, and $\varepsilon_{k,t}$ is the relevant error term. The logarithmic conditional variance of the market index is modeled by

$$log\left(\sigma_{k,t}^2\right) = \omega + \alpha \left[\frac{|\varepsilon_{k,t-1}|}{\sigma_{k,t-1}} - \frac{\sqrt{2}}{\pi}\right] + \beta \left(\frac{\varepsilon_{k,t-1}}{\sigma_{k,t-1}}\right) + \gamma log\left(\sigma_{k,t-1}^2\right) + \delta_i \Delta SENT_{k,i,t} \tag{4}$$

where $\sigma_{k,t}^2$ is the conditional variance of the returns of the $k$-type market index, $\varepsilon_{k,t-1}$ is the first-order autoregressive lag from (3), $\Delta SENT_{k,i,t}$ is the change in the $i$-type investor sentiment proxy related to the $k$-type index from $t$ to $t$-$1$, and $\omega, \alpha, \beta, \gamma, \delta_i$ are parameters to be estimated. $\omega$ is a constant and $\alpha$ represents the symmetric effect of the general autoregressive model. The $\beta$ coefficient captures the asymmetric effect, or the "leverage effect" of innovations on the volatility of the $k$-type market index returns. Therefore, if $< 0$ , negative innovations (i.e., bad economic news) generate higher volatility than positive innovations. $\gamma$ measures persistence in the conditional volatility irrespective of market shocks, i.e., when $\gamma$ is relatively large, the conditional market variance takes a long time to fade out.

---

[14] Bollerslev, Chou and Kroner (1992) explain that the logarithmic form of the EGARCH model avoids negative variance estimation, and hence relaxes estimation constraints imposed on the model, making it one of the primary motivations for using the EGARCH model.

We are interested in exploring how changes in investor sentiment proxies affect the conditional volatility of the stock and bond market returns, and thus our main interest is the $\delta_i$ coefficient, which captures the impact of a change in the $i$-type investor sentiment proxy on the conditional variance of the $k$-type market index returns. The null hypothesis is $H_0: \delta = 0$, while rejecting the null indicates that a change in the $i$-type sentiment measure affects the conditional volatility of $k$-type market index returns (See Lee *et al.* (2002) and Verma and Verma (2007)).

In light of the strong evidence of flight-to-quality and flight-to-liquidity in U.S. capital markets during the subprime crisis of 2008-2009 (Dick-Nielsen et al., 2012; Friewald et al., 2012), it would be reasonable to assume that investor sentiment effects might be substantially different during financial crises, as opposed to more stable periods. In order to control for the varying effect of sentiment in times of crisis, we follow Dick-Nielsen et al. (2012) and define three sub-periods of interest: (1) the period before the financial crisis (January 2000 – August 2009); (2) the financial crisis period (September 2008 – May 2009); and (3) after the financial crisis (June 2009 – March 2019).

Assuming that behavioral explanations generate, at least partially, a time-series momentum (Daniel et al., 1998), these indicators should have a positive correlation with investor sentiment (i.e., a positive momentum proxy for high investor sentiment). Thus, in times when market conditions are normal, we expect the coefficient of the change in SMMI and BMMI to be positive, indicating that higher momentum is followed by an increase in the volatility of the market returns. On the other hand, during a financial crisis, when capital constraints become binding and investor sentiment is generally low, we expect the coefficient of the change in SMMI and BMMI to be negative, due to the adverse effect of the change in investor sentiment on the conditional volatility. In a similar manner, since the PCR and VIX indicators are widely viewed as bearish indicators (Brown and Cliff, 2004), and hence have a negative correlation with sentiment, we expect the $\delta$ coefficient of the change in SMSI, SVIX, and BMSI to be positive, with a much higher magnitude during the financial crisis. The scale of the $\delta$ coefficient will be compared between the investor sentiment proxies and between the sub-periods in order to understand which of the sentiment indicators has the largest effect on the conditional volatility in the TASE.

## 5. Results

In the following sections we show sentiment effects through regressions and EGARCH (1,1) fitting model estimations for both the stock and bond markets in the TASE.

### 5.1. Stock market sentiment

#### 5.1.1. Regressions

Daily returns of TA-35 are used in the regression analysis in order to explore the effect of each market sentiment indicator on the stock market daily return variations. **Table 1** summarizes the two-stage regression results for the TA-35 index and its respective investor sentiment proxies: SMMI, SMSI and SVIX.

Consistent with the weak-form efficient market hypothesis, we find in Panel A that $R_{\text{TA}-35,t-1}$ is not significant, and the adjusted R-squared of the estimation of equation (1) is low (0.03). This means that variations in the TA-35 index cannot be explained by the index on the previous day. The squared residuals from the estimation of equation (1) may capture the variations in the market due to non-economic factors, e.g., investors' sentiment. We use the squared residuals in order to quantify the effect of the investor sentiment proxies on market volatility.

Panel B of Table 1 shows the estimation results of equation (2) for each investor sentiment proxy related to TA-35 index. The results show high statistical significance for SMMI and SVIX (prob.<0.01), with respective adjusted R-squares of 0.028 and 0.137, meaning that these two investor sentiment proxies explain daily market volatility. As expected, the coefficient of SMMI is negative (-0.053), implying that when the SMMI is positive and increasing, i.e. in times of a positive investor sentiment in the TASE, the variability of the TA-35 index returns declines. For the SVIX coefficient, we show a positive relation between the SVIX and squared residuals, implying that the variability of the TA-35 index increases when the SVIX increases, which is to be expected because the SVIX measures the volatility level in the market. The SMSI indicator, which is based on the ratio of the volume of put relative to call options, is not significant in the OLS regression test, and thus cannot explain variability in the TA-35 index. These results are consistent with the findings of Brown and Cliff (2004).

**Table 1: Two-stage regression results of TA-35 index volatility and its respective investor sentiment proxies**

We test for potential effects of the investor sentiment proxies on TA-35 index volatility by using OLS regression of the squared residuals obtained from equation (1) against each investor sentiment proxy. This table shows the regression results of the TA-35 daily returns against the lagged daily returns in Panel A, and summarizes regression results for each regression test in Panel B. For each regression, the table shows the number of observations (after adjustments), the regression coefficient, the standard error and the respective $t$-statistics and adjusted $R^2$. The data are the daily TA-35 index simple returns and the sample period is January 19, 2000 – March 18, 2019.

Panel A: Regression estimation results of the form $R_{TA35,t} = \beta_0 + \beta_1 R_{TA-35,t-1} + \varepsilon_{k,t}$,

| Variable | Coefficient | Std. Error | $t$-Statistic | Prob. | Observations | Model adjusted $R^2$ |
|---|---|---|---|---|---|---|
| Intercept | 0.03 | 0.017 | 1.730 | 0.083 | N=4,703 | 0.030 |
| $R_{TA-35,t-1}$ | 0.02 | 0.015 | 1.400 | 0.162 | | |

Panel B: Regression estimation results of the form:

$\left(R_{TA35,t} - \hat{R}_{TA35,t}\right)^2 = \beta_0 + \beta_1 SENT_{TA-35,i,t} + \varepsilon_{TA-35,t}$

| Variable | Coefficient | Std. Error | $t$-Statistic | Prob. | Observations | Model adjusted $R^2$ |
|---|---|---|---|---|---|---|
| Intercept | 1.566 | 0.053 | 29.759 | 0.000 | N=4,703 | 0.028 |
| SMMI | -0.053 | 0.004 | -11.652 | 0.000 | | |
| | | | | | | |
| Intercept | 0.674 | 0.400 | 1.686 | 0.092 | N=2,275 | 0.000 |
| SMSI | 0.113 | 0.387 | 0.292 | 0.770 | | |
| | | | | | | |
| Intercept | -1.424 | 0.113 | -12.606 | 0.000 | N=4,703 | 0.137 |
| SVIX | 0.178 | 0.006 | 27.286 | 0.000 | | |

### 5.1.2. EGARCH (1,1) Results

Following Uygur and Taş (2014), we model the conditional volatility of TA-35 market index returns as a function of the change of investor sentiment proxies. This allows us to explore whether a change in a given $i$-type investor sentiment proxy is related to a change in the conditional variance of the stock market, and how this effect differs between the three sample sub-periods (before, during and after the financial crisis of 2008). Table 2 summarizes the results of the EGARCH model on TA-35 daily returns as a function of the change in SMMI and SVIX.[15] The table reports the coefficients of $\omega, \alpha, \beta, \gamma, \delta$ and the respective standard errors of the EGARCH (1,1) model, estimated for the three different sub-periods.

The EGARCH coefficients show positive and highly significant $\alpha$ values in periods before and after the subprime crisis. These results are indicative of an ARCH effect in times of relatively stable markets, in which the current volatility of TASE returns is highly sensitive to the prior period's market events. In contrast, we find an insignificant $\alpha$ coefficient during the subprime crisis, indicating that there is no ARCH effect in times of financial distress. The results also show a negative and highly significant $\gamma$ coefficient in the period before the subprime crisis, indicating persistence of the conditional volatility during this period; however, we find an insignificant $\gamma$ for the period of the financial crisis, indicating that there is no persistence in the conditional volatility during this period.

---

[15] Note that the effect of the change in SMSI cannot be modeled in EGARCH, which requires a continuous data sample.

## Table 2: EGARCH (1,1) model results

We test the effect of the change in investor sentiment proxies on the conditional volatility of TASE market returns based on the EGARCH (1,1), with mean and variance equations according to equations (3) and (4) respectively. The table reports estimation coefficients and standard errors (in parenthesis), the model's adjusted $R^2$, AIC and Schwarz criterion of the EGARCH (1,1) model, estimated for the three sample sub-periods. The data are the daily TA-35 index simple returns and the sample periods vary by sample.

Panel A: Estimation results of the mean equation: $r_t = \mu + \varepsilon_t,$

| Variable | Before Crisis | Subprime Crisis | After Crisis |
|---|---|---|---|
| Intercept | 0.067 (0.0226)*** | .0960 (0.1563) | 0.0556 (0.0132)*** |

Panel B: Estimation results of the variance equation:

$$\log(\sigma_{TA35,t}^2) = \omega + \alpha\left[\left|\frac{\varepsilon_{TA35,t-1}}{\sigma_{TA35,t-1}}\right| - \frac{\sqrt{2}}{\pi}\right] + \beta\left(\frac{\varepsilon_{TA35,t-1}}{\sigma_{TA35,t-1}}\right) + \gamma\log(\sigma_{TA35,t-1}^2) + \delta_i\Delta SENT_{TA35,i,t}$$

| Variable | Before Crisis | Subprime Crisis | After Crisis |
|---|---|---|---|
| $\omega$ | 0.4164 (0.0646)*** | 1.990 (0.4773)*** | -0.7283 (0.0595)*** |
| $\alpha$ | 0.1759 (0.0425)*** | -0.1362 (0.2232) | 0.4509 (0.0487)*** |
| $\beta$ | -0.0555 (0.0295)* | -0.1482 (0.1438) | -0.0823 (0.0334)** |
| $\gamma$ | -0.2603 (0.0435)*** | -0.2536 (0.2351) | 0.0506 (0.0438) |
| $\Delta SMMI$ | -0.1501 (0.1232) | -0.7703 (0.4682)* | -0.4443 (0.1843)** |
| $\Delta SVIX$ | 0.8803 (0.0431)*** | 0.4743 (0.1322)*** | 1.3248 (0.0717)*** |
| Adjusted R-squared | -0.0003 | -0.0035 | -0.0009 |
| Akaike Info Criterion | 3.2618 | 4.4157 | 2.3391 |
| Schwarz Criterion | 3.2813 | 4.5593 | 2.3582 |
| Time Period | 01/20/2008 – 08/31/2008 | 09/01/2008 – 05/31/2009 | 06/01/2009 – 03/18/2019 |
| Number of Observations | 2,362 | 177 | 2,407 |

*, ** and *** indicate significance at the 10%, 5% and 1% level respectively

As expected, the results show negative and statistically significant $\beta$ coefficients in periods before (prob.<0.1) and after (prob.<0.05) the subprime crisis, and insignificant $\beta$ during the subprime crisis. This finding is indicative of a leverage effect only in times of normal market conditions. It implies that during normal market conditions there is a negative autocorrelation between past returns and future volatility, meaning that bad news or negative sentiment in the stock markets have a higher impact on the conditional variance of TA-35 index returns than a positive sentiment. This outcome is similar to the findings reported by Lee et al. (2002) and Verma and Verma (2007), who show a greater effect of bearish than bullish investors on the conditional volatility of the stock market returns. In contrast, we find that in times of financial distress, when market conditions are binding and asset prices drop dramatically, the leverage effect fades out.

Estimated coefficients of $\Delta SMMI$ show statistical significance during and after the subprime crisis (prob.<0.05) but not beforehand, possibly due to the technology crash of the year 2000. There is a negative coefficient for $\Delta SMMI$ of -0.44 in the period after the subprime crisis, suggesting an adverse effect of $SMMI$ on the conditional volatility of the local stock market's returns. That is, a negative shock in $SMMI$, which implies a negative investor sentiment and a bearish change in market returns, leads to an increase in the conditional volatility of the market, as stated by Lee et al. (2002). This implies that during normal market conditions the market momentum indicator may serve as a good proxy for investor sentiment in the TASE, since it has a large effect on the conditional volatility of market returns. We also observe the adverse effect of $\Delta SMMI$ and the conditional volatility of TA-35 returns during the subprime crisis, with a negative $\Delta SMMI$ coefficient of -0.77 (prob.<0.1). This result implies that in times of financial distress, a negative change in market momentum has a larger effect on the conditional volatility of stock returns than in times of normal market conditions. This result may be explained by the flight-to-liquidity effect in times of financial distress, where highly illiquid stocks tend to have greater volatility of returns, as documented by Acharya and Pedersen (2005).

As expected, the $\Delta SVIX$ has a positive and highly significant coefficient in all three sub-periods, suggesting that there is a large positive effect of the change in $SVIX$ on the conditional volatility of TA-35 index return. This means that an increase in the $SVIX$ sentiment indicator, pointing to a rise in future market risk expectations and a lower investor sentiment, significantly increases the conditional variance of TASE index returns. Since an increase in $SVIX$ is largely attributed to higher fear and stress in the market, and subsequently to a lower investor sentiment, this result is also consistent with the negative effects of sentiment on volatility reported by Verma and Verma (2007). Our results further show large differences in the $\Delta SVIX$ coefficient between the three sub-periods. We observe a coefficient of 0.47 during the subprime crisis, while we observe a coefficient of 0.88 and 1.32 before and after the subprime crisis, respectively. These values suggest that during normal periods, a temporary hike in volatility between *t* and *t-1* has a much greater effect on the conditional volatility of TA-35 returns than in times of financial distress.

## 5.2. Bond market sentiment

### 5.2.1. Regressions

Table 3 summarizes the two-stage regression results of the Tel-Bond-20 index with its respective investor sentiment proxies: BMMI, BSI and DRI. The results in Panel A show that the majority of the variation in the Tel-Bond-20 index returns is explained by the previous returns of the index, as evident by the highly significant value of $R_{Tel-Bond\,20,t-1}$ (prob. <0.01) in the estimation of equation (1).

**Table 3: Two-stage regression results of Tel-Bond-20 index volatility and its sentiment proxies**

We test for the potential effect of the investor sentiment proxies on Tel-Bond-20 index volatility by using OLS regression of the squared residuals obtained from equation (1) against each investor sentiment proxy. This table shows the regression results of the Tel-Bond-20 daily returns against the lagged daily returns (Panel A), and summarizes regression results for each regression test in Panel B. For each regression, the table shows the number of observations (after adjustments), the regression coefficient, the standard error and the respective t-statistics and adjusted $R^2$. The data are the daily Tel-Bond-20 index simple returns and the sample period is October 2, 2008 – March 18, 2019.

Panel A: Model estimation results:

$$R_{Tel-Bond\ 20,t} = \beta_0 + \beta_1 R_{Tel-Bond\ 20,t-1} + \varepsilon_{k,t},$$

| Variable | Coefficient | Std. Error | $t$-Statistic | Prob. | Number of Observations | Model adjusted $R^2$ |
|---|---|---|---|---|---|---|
| Intercept | 0.016 | 0.007 | 2.340 | 0.019 | | |
| $R_{Tel-Bond-20,t-1}$ | 0.191 | 0.018 | 10.607 | 0.000 | N=2,969 | 0.036 |

Panel B: Regressions estimation results:

$$\left(R_{Tel-Bond\ 20,t} - \hat{R}_{Tel-Bond\ 20,t}\right)^2 = \beta_0 + \beta_1 SENT_{Tel-Bond\ 20,i,t} + \varepsilon_{k,t}$$

| | | | | | | |
|---|---|---|---|---|---|---|
| Intercept | 0.225 | 0.023 | 9.574 | 0.000 | | |
| BMMI | -0.036 | 0.006 | -5.882 | 0.000 | N=2,722 | 0.012 |
| Intercept | -0.236 | 0.025 | -9.513 | 0.000 | | |
| BMSI | 0.089 | 0.004 | 20.476 | 0.000 | N=2,969 | 0.123 |
| Intercept | 0.0003 | 0.006 | 0.048 | 0.962 | | |
| DRI | 0.010 | 0.001 | 10.496 | 0.000 | N=1,854 | 0.056 |

Table 3 also shows estimation results of equation (2) for all investor sentiment proxies related to the Tel-Bond-20 index. We find that all three investor sentiment proxies are highly significant (Prob. <0.01), while the BMSI has a greater explanatory power than the BMMI and DRI indexes (with respective adjusted R-squared of 0.123, 0.012 and 0.055). As expected, the results show a negative estimated coefficient for the BMMI indicator (-0.036) and a positive estimated coefficient for the BMSI (0.089), implying that a positive momentum of investor sentiment in the bonds market, which is reflected by a higher BMMI and a lower BMSI, significantly lowers the variability of the Tel-Bond-20 index. For the DRI index, a measure of the risk of default in the bonds market, the estimation results show a positive coefficient of 0.010, implying that when the DRI index increases, the variability of the Tel-Bond-20 index increases. This result may be explained by the *flight-to-quality* effect, where the yield to maturities of corporate bonds rise dramatically as a result of shifts in investors' holdings towards the high quality assets. Hence, the DRI index is closely related to the variability of the Tel-Bond-20 market returns.

### 5.2.2. EGARCH (1,1) Results

We study the effect of a change in bond market investor sentiment proxies on the conditional variance of the Tel-Bond-20 index returns. Table 4 summarizes the results of the EGARCH model on the Tel-Bond-20 daily returns as a function of the change in BMMI and BMSI.[16] The table summarizes the estimated coefficients of $\omega, \alpha, \beta, \gamma, \delta$ and the respective standard errors of the EGARCH (1,1) model, estimated for the three different sub-periods.

Table 4 shows a positive and highly significant $\alpha$ (prob.<0.01) for all estimated sub-periods, suggesting that the volatility of Tel-Bonds 20's returns is highly sensitive to market events.

**Table 4: EGARCH (1,1) model results**

We test the impact that a change in sentiment proxies has on the conditional volatility of Tel-Bond-20 returns with an EGARCH (1,1) model, as specified in equations (3) and (4). The table shows estimation coefficients and standard errors, and the model's adjusted $R^2$, AIC and Schwarz criterion. The data are the daily Tel-Bond-20 index simple returns and the sample period is January 19, 2000 – March 18, 2019.

---

[16] Note that the effect of the change in DRI cannot be modeled in EGARCH, which requires a continuous data sample.

Panel A: Estimation results of the mean equation:

$$r_t = \mu + \varepsilon_t,$$

| Variable | Before Crisis | Subprime Crisis | After Crisis |
|---|---|---|---|
| Intercept | 0.085 (0.0175)*** | -0.030 (0.0576) | 0.022 (0.0034)*** |

Panel B: Estimation results of the variance equation:

$$\log(\sigma_{TA35,t}^2) = \omega + \alpha\left[\left|\frac{\varepsilon_{TA35,t-1}}{\sigma_{TA35,t-1}}\right| - \frac{\sqrt{2}}{\pi}\right] + \beta\left(\frac{\varepsilon_{TA35,t-1}}{\sigma_{TA35,t-1}}\right) + \gamma\log(\sigma_{TA35,t-1}^2) + \delta_i\Delta SENT_{TA35,i,t}$$

| Variable | Before Crisis | Subprime Crisis | After Crisis |
|---|---|---|---|
| $\omega$ | -0.8406 (0.1039)*** | -0.7873 (0.2723)*** | -0.1638 (0.0273)*** |
| $\alpha$ | 0.9049 (0.1543)*** | 0.6105 (0.2465)*** | 0.1445 (0.0251)*** |
| $\beta$ | -0.1281 (0.0887) | -0.0287 (0.1296) | -0.0611 (0.0132)*** |
| $\gamma$ | 0.8184 (0.0390)*** | 0.0922 (0.1520) | 0.9841 (0.0040)*** |
| $\delta\Delta BMMI$ | 0.1896 (0.3261) | -0.8076 (0.6280) | 0.2532 (0.0826)*** |
| $\delta\Delta BMSI$ | 0.3778 (0.0864)*** | 0.7957 (0.1675)*** | 0.2563 (0.0631)*** |
| Adjusted R-squared | -0.0033 | -0.0005 | -0.0005 |
| Akaike Info Criterion | 1.6347 | 2.6151 | -0.4017 |
| Schwarz Criterion | 1.7178 | 2.7586 | -0.3824 |
| Time Period | 02/11/2008 – 08/31/2009 | 09/01/2008 – 05/31/2009 | 06/01/2009 – 03/18/2019 |
| Number of Observations | 379 | 177 | 2,407 |

*, ** and *** indicate significance at the 10%, 5% and 1% level respectively

As expected, in the period after the subprime crisis, the table reports a negative and statistically significant $\beta$ coefficient (prob.<0.01), an indication of a leverage effect and a negative correlation between past returns and future volatility of the Tel-Bond-20 index returns. Hence, in normal times, a negative shock in the bonds market has greater impact on the conditional variance of Tel-Bond-20 returns than a positive shock has. This finding is similar to the findings reported by Nayak (2010), who shows that sentiment-driven mispricing and systematic reversal trends in the bonds market are very similar to those for the stocks market. Nonetheless, in the periods before and during the subprime crisis, our results show an insignificant $\beta$ coefficient, suggesting that there is no leverage effect in the bonds market in times of financial distress. This outcome may be explained by the flight-to-quality effect before and during the subprime crisis, in which many investors toss their holdings in corporate bonds towards higher quality bonds, predominantly government bonds. Thus, negative shocks, while significantly outnumbering positive shocks, exhibit an insignificant leverage effect.

Our results show differences in the $\gamma$ coefficients between subperiods, which capture the persistence of conditional volatility irrespective of market shocks. As expected, the $\gamma$ coefficient is found to be positive and highly significant in the periods before and after the subprime crisis, implying that during these periods, shocks to the conditional variance of Tel-Bond-20 will be highly persistent, i.e., a large noisy signal (positive or negative) will lead future variance to be high. In contrast, during the subprime crisis, persistence of the conditional variance fades out, implying that a temporary shock cannot affect future variance of bond market returns.

Regarding the coefficients of changes of investor sentiment proxies ($\delta$), $\Delta BMMI$ has a positive and highly significant coefficient of 0.2532 in the normal period following the subprime crisis, which implies a positive effect on the conditional volatility of Tel-Bond-20 index returns. In the period before and during the subprime crisis, the $\Delta BMMI$ coefficient is not significant. These results suggest that during a normal period, where market conditions are normal and the overall sentiment is positive, a decrease in the BMMI, which implies a decline in momentum, is associated with a decline in the conditional volatility. While the coefficient during the financial crisis is negative, and thus suggestive of a contrary effect (when sentiment is low, a decrease in BMMI increases conditional volatility), this coefficient is not significant, possibly due to much noise.

Concerning the BMSI index, our results show a positive and highly statistically significant $\Delta BMSI$ coefficient in all periods, implying that a positive change in the BMSI increases the conditional volatility of the Tel-Bond-20 index returns. Hence, an increase in the BMSI, which reflects higher market risk expectations and a decrease in sentiment, is significantly related to an increase in the conditional variance of Tel-Bond-20 index returns. As expected, the coefficient of $\Delta BMSI$ during the subprime is much higher than the estimated coefficients found before and after the subprime crisis (0.79, 0.37 and 0.25 respectively).

In summary, our results show that momentum-based indicators can explain returns volatility in the Israeli stock and bonds market. For the stock market, we show that a positive change in momentum (i.e., positive change in SMMI) also increases the conditional volatility of TA-35 returns, while a positive change in BMMI reduces the conditional volatility of Tel-Bond-20 index returns. Yet, we find that the change in momentum affects the conditional volatility of Tel-Bond-20 index only in times of normal market conditions. For the volatility-based indicators, we find that a change in these indicators also increases the conditional volatility of the returns in both the TA-35 and Tel-Bond-20 indexes and in all sub-periods; however they differ in magnitude. We find that in times of normal market conditions, a positive change in the implied volatility-based SVIX indicator typically has a larger effect on the conditional volatility of the stock market returns than in times of financial distress. In contrast, we observe a lower effect of $\Delta BMSI$ on the conditional volatility of the Tel-Bond-20 index returns during the periods before and after the subprime crisis.

## 6.  Conclusions

An extensive body of literature shows that noise traders may affect financial asset prices. Because rational investors should not trade on noise, behavioral trades due to sentiment may help explain those trades. We use several proxies of market sentiment indicators in order to study the effect that noise traders may have on the conditional volatility of the stock and bond markets' index returns in the TASE. Our test of bond market sentiment appears to be the first to measure the sentiment of retail investors, as they are highly active in the Israeli limit order book market, as opposed to OTC bond traders in most of the developed exchanges. Additionally, given our unique dataset of retail investors in corporate bonds, our paper is the first to explore bond market sentiment on future, conditional volatility of corporate bond returns.

Using an EGARCH model on the TA-35 and Tel-Bond-20 index returns, we show that a change in a market sentiment proxy, which reflects a change in risk expectations and investor sentiment, largely explains movements in the conditional volatility of both stock and bond market returns. More specifically, our results show that momentum-based indicators and volatility-based indicators are closely related to bond and stock market return volatility in the TASE. We also find that the index that captures risk of default in the bond market explains the volatility of the corporate bond index.

Given the low sentiment that was measured during the financial crisis of 2008, we find that the change in volatility-based sentiment proxies has a larger effect on the conditional volatility of corporate bond index returns than in times of normal market conditions, a pattern that is reversed in the stock index returns.